\documentstyle[aps,prl,epsfig,floats]{revtex} 
\draft
 
\begin{document} 

\twocolumn[\hsize\textwidth\columnwidth\hsize\csname
@twocolumnfalse\endcsname

\title{Neutrinos as a signature of ultrahigh energy photons at high red
shift} 
 
\author{Marieke Postma} 
\address{Department of Physics and Astronomy, UCLA, Los Angeles, 
CA 90095-1547
}           
 
\date{\today}

\maketitle
 
\begin{abstract}
Sources of ultrahigh energy photons operating at high red shift
produce a diffuse background of neutrinos.  At high red shift, when
the cosmic microwave background radiation has a higher temperature, an
electromagnetic cascade originated by an energetic photon can generate
neutrinos via muon and pion production and decay.  We have calculated
numerically the neutrino spectrum produced by various photon sources.
A distinctive feature of the produced flux is a ``bump'' in the
spectrum at neutrino energies $E \sim 10^{17}$eV.  The produced flux
is largest for $m=3$ sources ({\it e.g} necklaces), with $E^2 J(E)
\sim 10\, {\rm eV} {\rm cm}^{-2} {\rm s}^{-1} {\rm sr}^{-1}$ at these
energies.  Detection of such neutrinos can help understand the origin
of ultrahigh energy cosmic rays.
\end{abstract} 
 
\pacs{PACS numbers: 98.70.Sa, 95.85.Ry, 98.70.Vc 
\hspace{1.0cm}  UCLA/01/TEP/2
}

\vskip2.0pc]

\renewcommand{\thefootnote}{\arabic{footnote}}
\setcounter{footnote}{0}

\section{Introduction}

In the near future several experiments will become sensitive to small
fluxes of very high energy neutrinos.  One may hope to observe
neutrinos from active galactic nuclei (AGN), gamma ray bursts, other
astrophysical sources, and, possibly, from yet undiscovered objects
that have emerged from particle theory, {\em e.g.}, topological
defects (TD).  To learn new physics from the future observations, it
is important to identify signatures of certain sources that may
contribute to diffuse neutrino flux.  In this paper we detail the
description of one such signature~\cite{kp}, namely the imprint of
ultrahigh energy photons from high red shift on the neutrino spectrum.
At high red shift, when the cosmic microwave background radiation
(CMBR) has a higher temperature, an electromagnetic cascade originated
by an ultrahigh energy photon can generate neutrinos via muon and pion
production and decay.  The present-day energies of such neutrinos are
about $10^{17}$~eV.  Detection of a ``bump'' in the neutrino spectrum
at $10^{17}$~eV can indicate the presence of ultrahigh energy photon
sources at high red shift.

The question of ultrahigh energy photon sources was brought in sharp
focus by recent observations of cosmic rays~\cite{data} with energies
beyond the Greisen-Zatsepin-Kuzmin (GZK) cutoff~\cite{gzk}.  The
origin of such cosmic rays remains an outstanding
puzzle~\cite{reviews}. Most of the proposed explanations can be
categorized in two main classes.  In the {\em bottom-up/acceleration}
scenarios, charged particles are accelerated to ultrahigh energies in
giant astrophysical ``accelerators'', such as active galactic nuclei
and radio galaxies~\cite{conv,mpr}.  In the {\em top-down/decay}
scenario on the other hand, massive objects such as topological
defects~\cite{TD_review,TD,TD_ber,TD_nu} and superheavy relic
particles~\cite{particles,gk2,kt} decay, emitting mainly ultrahigh
energy photons.  Topological defects and relic particles can exist at
much higher red shift than the astrophysical candidate sources, which
are formed only after the onset of galaxy formation.

To understand the origin of the ultrahigh energy cosmic rays (UHECR),
it is crucial to distinguish between these two different scenarios.
In~\cite{kp} we showed that sources of ultrahigh energy photons
operating at high red shift produce a diffuse background of neutrinos
with energies $E_\nu \sim 10^{17}$eV.  Our rough estimates indicated
that for some photon sources, {\it e.g.} necklaces, this neutrino flux
is large enough to be detectable in the near future.  It is therefore
useful to get a better estimate of the produced neutrino flux.  To
this end we have done a numerical calculation.  In the present paper
we will present the results.

Generation of ultrahigh energy neutrinos has been
studied~\cite{TD,TD_ber,TD_nu,conv,mpr,hs,stecker,ysl,pj,ps,whitepaper}
for various sources at small red shift, for which muon pair production
can be neglected.  However, a substantial flux of neutrinos could be
produced at earlier times, when the propagation of photons was
different from that in the present universe because the intergalactic
magnetic field was weaker, the density of radio background was lower,
and the cosmic microwave background density and temperature were
higher.  At red shift $z$ the temperature of the cosmic microwave
background radiation increases by a factor $(1+z)$.  Because of this,
at high red shift ultrahigh energy photons and electrons can produce
muons through interactions with the CMB photons. Muons decay into
neutrinos.  This is in sharp contrast with the $z \lesssim 1$ case,
where the photons do not produce neutrinos as they lose energy mainly
by scattering off the radio background and in the subsequent
electromagnetic cascade~\cite{berezinsky,reviews}.

In section~\ref{production} we will describe in detail this neutrino
production mechanism.  We will give an estimate for the produced flux
of neutrinos in section~\ref{estimates}; the numerical results are
presented in section~\ref{numerics}.  Our discussion applies to any
source of photons active at high red shift.  The latter requirement
excludes most astrophysical sources~\cite{conv}.  Topological defects
and decaying relic particles on the other hand, could operate at much
higher red shifts.  These sources are expected to produce photons with
energies as high as $10^{20}$eV.  Their presence need not be connected
to the GZK puzzle.  Superconducting strings and other rapidly evolving
topological defects, for example, are ruled out as the origin of
observed ultrahigh energy cosmic rays, because they would produce too
large a flux of secondary, low-energy photons.  However, despite this
constraint on their present density, such sources could have existed
in the early universe and could produce a detectable flux of
neutrinos.

\section{Neutrino production}
\label{production}

The propagation of energetic photons and electrons is governed by
their interactions with the cosmic background radiation.  At small red
shift they lose energy mainly through interactions with the radio
background. The main interactions involved in the electromagnetic
cascade are pair production $\gamma \gamma \to e^+e^-$ and inverse
Compton scattering $e \gamma \to e \gamma$.  The cross section for
pair production is peaked near the threshold, where $E_\gamma \epsilon
\simeq m_e^2$.  Therefore, for ultrahigh energy (UHE) photons the most
effective targets are background photons with energy $\epsilon \approx
m_e^2/E_\gamma \lesssim 10^{-6} {\rm eV}$, i.e., radio photons. The
radio background is generated by normal and radio galaxies.  Its
present density~\cite{biermann} is higher than that of CMB photons in
the same energy range.  As a result, the mean free path for energetic
photons and electrons is determined by the radio background, and at
small red shift the main source of energy loss for UHE photons is
electromagnetic cascade generated by interactions with radio photons.
At red shift $z$, however, the density of CMB photons is higher by a
factor $(1+z)^3$, while the density of radio background is either
constant or, more likely, lower.  Some models of cosmological
evolution of radio sources~\cite{condon} predict a sharp drop in the
density of radio background at red shift $z \gtrsim 2$.  More recent
observations~\cite{recent} indicate that the decrease of radio
background at $z>2$ is slow.  In any case, at high red shift one
expects the CMB to set the mean free path and thus become the dominant
source of energy loss for photons, because of the $(1+z)^3$ increase
in the density.  Let $z_{_{\rm R}}$ be the red shift at which
scattering off radio background can be neglected with respect to
scattering of CMBR.  Based on the analyses of
Refs.~\cite{condon,recent}, we take $z_{_{\rm R}} \sim 5$.

Another source of energy loss is the synchrotron radiation emitted by
the cascade electrons in the intergalactic magnetic field (IGMF).
This is an important effect at small red shift, but it becomes less
significant at earlier times when the IGMF is weak.  For $z>z_{_{\rm
M}} \sim 5$ synchrotron losses are small compared to energy losses
resulting from interactions with the CMB radiation. We will use the
value $z_{\rm min}=\max (z_{_{\rm R}}, z_{_{\rm M}}) \approx 5$ in what
follows.  As discussed in section~\ref{numerics}, a higher value of
$z_{\rm min}$, even as high as 10, would not make a big difference in
the flux of the signature neutrinos.

Not only the density of the CMB increases with red shift, so does its
temperature: $T_{_{CMB}}(z) = 2.7 (1+z){\rm K}$.  The center of mass
energy in scatterings of UHE electrons and photons with CMB photons
increases, and new, neutrino producing interactions become possible.
At red shift $z>z_{\rm min}$, where radio background and the
intergalactic magnetic field can be neglected, photons and electrons
scattering off the CMB can produce muons through the processes $\gamma
\gamma_{_{CMB}} \rightarrow \mu^+ \mu^-$ and $e \gamma_{_{CMB}}
\rightarrow e \mu^+ \mu^- $. Muons decay into neutrinos: $\mu
\rightarrow e \nu_e \nu_{\mu}$.  The threshold for these muon
producing interactions is $\sqrt{s}> 2 m_\mu=0.21$GeV, or
\begin{equation} 
E_{\gamma,e} > E_{\rm th}(z)=\frac{ 10^{20}{\rm eV}}{1+z} 
\label{threshold} 
\end{equation} 

But will muons, and thus neutrinos, indeed be produced? To answer this
question one has to look closely at the propagation of UHE photons at
$z \gtrsim 5$.  Photons will scatter off the CMBR; for photon energies
above the threshold for muon pair production, the reactions $\gamma
\gamma_{_{CMB}}\rightarrow e^+ e^-$, $\; \gamma
\gamma_{_{CMB}}\rightarrow e^+ e^- e^+ e^-$ and $\gamma
\gamma_{_{CMB}}\rightarrow\mu^+ \mu^-$ are possible.  For $\sqrt{s}>
2m_{\pi^{\pm}} = 0.28$GeV charged pion production may also occur.
Accelerator experiments show that the pion cross section is small
compared to the cross sections for muon production at the energies of
interest~\cite{pi}. We will, therefore, neglect it.

The cross section for electron pair production (PP) decreases with
increasing photon energy
\begin{equation}
  \sigma_{\rm PP} = \frac{8 \pi \alpha^2}{s} \ln \frac{s}{m_e^2} 
  \qquad (s \gg m_e^2), 
\label{sig_PP}
\end{equation}
while the cross section for double pair production (DPP), a higher
order QED process, quickly approaches its asymptotic value~\cite{dpp}
\begin{equation}
\sigma_{_{\rm DPP}} \simeq  \frac{172 \alpha^4}{36 \pi m_e^2} 
	\simeq 6.45 \mu{\rm barn} \qquad (s \gg m_e^2). 
\label{dpp} 
\end{equation}
PP is the dominant reaction for photon energies $E_\gamma \lesssim 5
\times 10^{20} {\rm eV} /(1+z)$.  Since the energies of the two
interacting photons are vastly different, either the electron or the
positron from PP has energy close to that of the initial photon.  At
higher photon energies, DPP becomes more important.  One or more
energetic electrons are produced in this reaction.  Muon production is
suppressed at all energies, its cross section being smaller than that
for electron pair production by a factor $2$ at threshold to a factor
$10$ at higher energies. Thus after an initial $\gamma
\gamma_{_{CMB}}$ reaction, there is a small chance muons are produced,
but most likely one ends up with one or more UHE electrons.

These electrons continue to scatter off the CMBR.  At lower energies,
inverse Compton scattering, $e \gamma_{_{CMB}}\rightarrow e \gamma$,
converts high-energy electrons into high-energy
photons~\cite{reviews}.  However, at energies above the muon
threshold, higher order processes, such as triplet production (TPP) $e
\gamma_{_{CMB}}\rightarrow e\, e^+ e^-$ and electron muon-pair
production (MPP) $e \gamma_{_{CMB}}\rightarrow e\, \mu^+ \mu^-$,
dominate.  For higher energies charged pion production may also occur,
but we expect it to be suppressed by the same hadronic physics that
suppressed pion production in the photon-photon reaction, and we will
neglect it.  For center of mass energies $s \gg m_e^2$, the
inelasticity $\eta $ for TPP is very small~\cite{reviews,tpp,tpp2}:
\begin{equation}
\eta \simeq 1.768 (s/m_e^2)^{-3/4} < 10^{-3}.
\label{eta_tpp}
\end{equation}
Hence, the energy attenuation length $\lambda_{\rm eff}$ is much
larger than the TPP interaction length: $\lambda_{_{\rm TPP}} \simeq
\eta \lambda_{\rm eff}$.  One of the electrons produced through TPP
carries almost all ($1-\eta$) of the incoming electron's energy.  It
can interact once again with the CMBR. As a result, the leading
electron can scatter many times before losing a considerable amount of
energy.  Each time it scatters with a CMB photon there is another
chance to produce a muon pair, until the electron energy decreases
below threshold for muon production.  To determine whether muons are
produced, one must compare the {\em energy attenuation} length for
triplet production with the {\em interaction length} for muon-pair
production.  The interaction length is given by $ \lambda^{-1} \simeq
\left< n_{_{CMB}} \right> v \sigma$, and thus the ratio is
$R=\lambda_{\rm eff}/\lambda_{_{\rm MPP}} \simeq \sigma_{_{\rm
MPP}}/(\eta \sigma_{_{\rm TPP}})$. For $s \gg m_e^2$ the cross section
for TPP is~\cite{tpp,tpp2}
\begin{equation}  
\sigma_{_{TPP}} \simeq \frac{\alpha^3}{m_e^2} \left( \frac{28}{9} {\rm
ln}\frac{s}{m_e^2} -\frac{218}{27} \right).
\label{sig_TPP}
\end{equation}
The MPP cross section in the energy range just above the threshold $5
m_\mu^2<s<20 m_\mu^2$ is of the order of $0.1-1{\rm mb}$, and the
ratio in this range is $R \sim 100$.
 
Since $\lambda_{\rm eff} \gg \lambda_{_{\rm MPP}}$, in the absence of
dense radio background and intergalactic magnetic fields, all
electrons with $E \gtrsim E_{\rm th}$ pair-produce muons before their
energy is reduced by the cascade. For muon production close to the
threshold, each muon carries on average $1/4$ of the incoming
electron's energy~\cite{tpp2}. Muons decay before they can interact
with the photon background.  Each energetic muon produces two
neutrinos and an electron.  The electron produced alongside the muon
pair gets half or more of the incoming electron's energy; it can
interact again with the CMBR to produce muons. This process can repeat
until the energy of the regenerated electron decreases below the
threshold for muon pair production.

\section{Flux estimates}
\label{estimates}

Let us assume that the time dependence of the rate of photon emission
can be parametrized as
\begin{equation}
\dot{n}_{\gamma}(E_\gamma, t) = \dot{n}_{\gamma,0}(E) \left(
 \frac{t}{t_0} \right)^{-m}.
\label{z_dep}
\end{equation}
The subscript $0$ denotes the present-day value of the parameter.  The
value of $m$ depends on the source under consideration.  For long
lived relic particles and topological defects, the rate of photon
emission is indeed (approximately) of the above form, with $m=0$ for
decaying relic particles, $m=3$ for strings, monopolonium and
necklaces, and $m \ge 4$ for superconducting strings.  Monopolonium is
a bound monopole anti-monopole pair.  Necklaces are networks of
strings and monopoles, with two monopoles attached to each string: the
monopoles are the beads in a necklace of strings.

All photons produced at red shifts $z_{\rm min} \leq z \leq z_{\rm
max}$ contribute to the present neutrino flux.  Here $z_{\rm min} \sim
5$ is the minimum red shift for which radio background and magnetic
fields are negligible, and $z_{\rm max}$ is the maximum red shift for
which the universe is transparent to UHE neutrinos. The value of
$z_{\rm max}$ is determined by the neutrino interactions with the
relic neutrino background.  The absorption red shift for neutrinos
with energy $\sim 10^{17}$eV is $z_{\rm max}\sim 3 \times
10^3$~\cite{ggs}.  The neutrino flux is:
\begin{eqnarray} \label{nu.z} 
\phi_\nu & = & \xi \int^{z_{\rm max}}_{z_{\rm min}} {\rm d}t \,
\dot{n}_{\gamma}(z) \ (1+z)^{-4} \\ & = & \xi \frac{3}{-2a} t_0
\dot{n}_{\gamma,0}(E>E_{\rm th}) \, \, [(1+z_{\rm min})^a - (1+z_{\rm
max})^a], \nonumber
\end{eqnarray}  
where $a=(3m-11)/2$, and $\xi$ is the number of neutrinos produced per
UHE photon.  An UHE photon produces an UHE electron.  If energetic
enough, this electron produces a muon pair, and four neutrinos are
generated.  However, for initial electron energies just above
threshold, it may often occur that repeated TPP reactions decreases
the energy below threshold before any muons are produced.  Thus for
photons with energies close to threshold for muon production $\xi <
4$.  On the other hand, for higher photon energies one can have $\xi >
4$.  This is because higher energetic photons are more likely to
produce a double pair of electrons (DPP), {\it i.e.}, produce more
than one UHE electron.  Furthermore, the electron produced alongside
the muon pair in MPP may be energetic enough for a second round of
muon pair-production.  Our numerical calculation (see
section~\ref{numerics}) gives $\xi = 0.6$ for $E_\gamma (1+z) = 1
\times 10^{20}{\rm eV}$, $\;\xi = 4.0$ for $E_\gamma (1+z) = 2 \times
10^{20}{\rm eV}$, and $\xi = 7.7$ for $E_\gamma (1+z) = 3 \times
10^{20}{\rm eV}$.  We will take $\xi \sim 4$ in our estimate.

Rapidly evolving sources with $m \ge 4$, such as superconducting
strings, are ruled out as the origin of the observed ultrahigh energy
cosmic rays.  The strongest constraint on the present density of such
sources comes from measurements of the diffuse $\gamma$-background
below $100 \, {\rm GeV}$~\cite{reviews,TD_ber,egret}.  UHE photons
lose energy in the electromagnetic cascade involving PP and ICS.  When
the photon energy drops below PP threshold, the photon attenuation
length becomes small, and for $z \lesssim 10^3$ the universe is
transparent to these low energetic photons.  The flux of the cascade
photons in the energy range $10 \, {\rm MeV} < E <100 \, {\rm GeV}$
must be lower than the flux measured by the Energetic Gamma Ray
Experiment Telescope (EGRET): $\omega_{\rm cas} \le 2.6 \times 10^{-6}
{\rm eV} {\rm cm}^3$. Although the present density of $m \ge 4$
sources is constrained, it is conceivable that in the early universe
they may have existed in large enough number to produce an appreciable
neutrino flux.

For $m < 11/3$, $a < 0$, and, according to~eq.~(\ref{nu.z}), the
photon sources at smaller red shift, $z \sim z_{\rm min} \approx 5$,
are the most important.  But this is only true for the contribution
from the most energetic photons, those with energies above the
threshold at small red shift, as given by eq.~(\ref{threshold}). As
$z$ increases, the threshold energy for muon production decreases, and
lower energetic photons can also produce muons.  Moreover, since the
photon spectrum is a falling function of energy, low energy photons
are more abundant; they all contribute at large red shift.  Which
effect wins, which red shift is favored, depends on both the evolution
index $m$ and the shape of the photon spectrum.  For a photon spectrum
of the form $\dot{n}_\gamma \propto E_\gamma^{-\beta}$ however, the
red shift dependence of the flux simplifies.  Using that all photons
with energy $E>E_{\rm th}$ produce muons, the neutrino flux then
becomes
\begin{equation}
\phi_{\nu} = \xi \dot{n}_0 \frac{3}{2 |\beta - 1|} (10^{20} {\rm
eV})^{1-\beta} \,C,
\label{nu.beta}
\end{equation} 
with 	
\begin{equation}
C = \left\{ 
\begin{array}{ll}
 	b^{-1} [
	(1 + z_{\rm max} )^{b} - (1 - z_{\rm min} )^{b} ],
	&  \quad b \neq 0, \\ 
 	\ln (z_{\rm max}/z_{\rm min}),   
	& \quad b = 0.\\
\end{array} \right . 
\end{equation}
Here $\dot{n}_0$ is an overall normalization constant and $b = (2\beta
+ 3 m -13)/2$.  For $b < 0 $ small red shifts are most important,
whereas for $b \ge 0 $ large $z$ dominates.

Nevertheless, the most energetic neutrinos are all produced at small
red shift.  We will estimate the flux of these high energy neutrinos,
leaving the total flux for the numerical calculation. The highest
energy neutrinos are produced at red shift $z \lesssim 10$.  We will
assume that the photon spectrum is of the form $\dot{n}_\gamma =
\dot{n}_0 E_\gamma^{-\beta} (t/t_0)^m$.  If the photon sources are the
origin of the UHECR today one can use the observed UHECR flux to fix
the normalization constant $\dot{n}_0$.  This gives
\begin{equation}
\phi_\nu = \phi_{_{\rm CR}} \xi \frac{(z_{\rm min})^b - (z_{\rm
	max})^b} {1 - (z_{_{\rm GZk},\gamma})^b},
\label{norm}
\end{equation} 
where we have used that only photons emitted at red shift $0 < z <
z_{_{GZK},\gamma}$ contribute to the observed UHECR flux.  Photons
with energies $E_\gamma \sim 10^{20} {\rm eV}$ have an energy
attenuation length of $\sim 10 {\rm Mpc}$~\cite{reviews} in the
present universe; this corresponds to red shift $z_{_{GZK},\gamma} =
0.002$.  The observed UHECR flux of particles with $E > E_{_{\rm GZK}}
\sim 5 \times 10^{19} {\rm eV}$ is $\phi_{_{\rm CR}} \sim 10^{-19}
{\rm cm}^{-2} {\rm s}^{-1} {\rm sr}^{-1}$.  Taking $\beta = 2$, in
agreement with UHECR observations, and $z_{\rm max} = 10$ we then
obtain for the flux of energetic neutrinos:
\begin{equation}
\phi_{\nu} \sim \left \{
\begin{array}{llll}
  7 \times 10^{-20}\, {\rm cm}^{-2} {\rm s}^{-1} {\rm sr}^{-1}, 
	& \qquad m=0, \\
  1 \times 10^{-18}\, {\rm cm}^{-2} {\rm s}^{-1} {\rm sr}^{-1}, 
	& \qquad m=1, \\
  3 \times 10^{-17}\, {\rm cm}^{-2} {\rm s}^{-1} {\rm sr}^{-1}, 
	& \qquad m=2, \\
  7 \times 10^{-16}\, {\rm cm}^{-2} {\rm s}^{-1} {\rm sr}^{-1}, 
	& \qquad m=3. \\
\end{array} \right. 
\label{flux}
\end{equation}
A value of $\beta = 1.5$ or $\beta = 2.5$ changes the produced flux by
about a factor $2$. The above results are valid for photon sources
that are uniformly distributed throughout the universe. Sources that
cluster in galactic halos ({\it e.g.} supermassive particles) produce
a lower flux, since the clustering enhances their contribution to the
observed UHECR flux.  To get the neutrino flux for photon sources with
clustering properties, one has to divide the results in
eq.~(\ref{flux}) by the over-density of these sources in our galaxy.

The energy of the produced neutrinos at red shift $z$ is $E_\nu (z)
\sim E_\mu/3$.  It is then further red shifted by a factor
$(1+z)^{-1}$.  Assuming a falling photon spectrum, we expect most of
neutrinos to come from photons near the threshold,
eq.~(\ref{threshold}). We estimate the energy of the most energetic
neutrinos after the red shift $E_\nu \sim 10^{17}$eV.

The largest flux of neutrinos are produced by $m=3$ sources.  Examples
of such sources are necklaces and monopolonium. In general,
topological defects emit both UHE protons and photons.  When the
protons, and not the photons, are responsible for the UHECR the
neutrino flux can no longer be normalized as in
eq.~({\ref{norm}). Therefore, whenever we state results for necklaces
or monopolonium, we will use the differential photon flux $J(E)$
calculated in~\cite{TD_ber,pj,ps} for normalization instead.  For
these sources, $\dot{n}_{\gamma,0} (E>E_{\rm min}) \sim L^{-1}
\int_{E_{\rm min}} {\rm d}E \, J(E) $, with $L$ the length scale from
which the photons are collected.  Because the photon flux is a sharply
falling function of energy, $\, \dot{n}_{\gamma,0}$ is dominated by
photons with energies $E\sim E_{\rm min}$. Monopolonium clusters in
galaxies and has $L \sim L_{\rm gal} \sim 100{\rm kpc}$, the size of
our galaxy.  Their over-density in the galaxy is $\sim 2 \times
10^{5}$~\cite{TD_ber}.  Necklaces on the other hand are distributed
uniformly throughout the universe; for them $L = L_\gamma \sim 5$Mpc,
the photon absorption length at these energies.  We then obtain for
the flux of neutrino produced at low red shift:
\begin{equation}
\phi_{\nu} \sim \left \{
\begin{array}{ll}
  10^{-18}{\rm cm}^{-2} {\rm
  s}^{-1} {\rm sr}^{-1},  & \quad {\rm monopolonium} \, (m=3) , \\  
10^{-16}{\rm cm}^{-2} {\rm s}^{-1} {\rm
  sr}^{-1}, &\quad {\rm necklaces}\, (m=3). \\
\end{array} \right. 
\label{flux_td}
\end{equation}

\section{Numerical results}
\label{numerics}

\begin{figure}[t]
\centering
\hspace*{-5.5mm}
\leavevmode\epsfysize=6cm \epsfbox{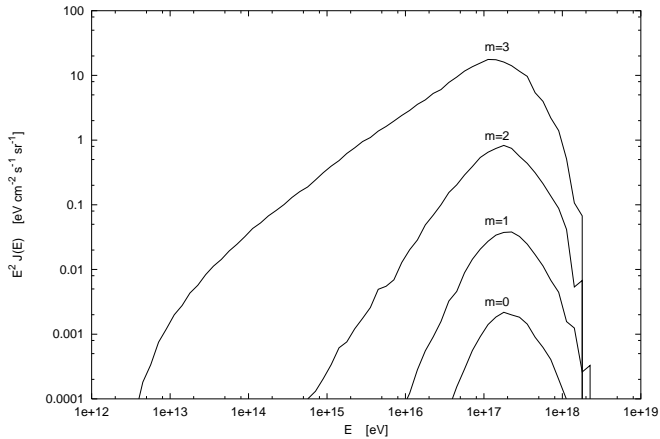}\\[3mm]
\caption[fig. 1]{\label{fig1} Differential neutrino flux $E^2 J(E)$
produced by sources with different evolution indices: $m = 0, \; 1, \;
2, \; 3$ and $4$. For all plots $\beta = 2$, $z_{\rm max} = 10^3$, and
$z_{\rm min} = 5$.}
\end{figure}

We have calculated numerically the neutrino flux produced by photon
sources at high red shift.  According to our estimate,
eq.~(\ref{flux}), $m=3$ sources produce the largest flux of neutrinos,
and we will concentrate on them. In particular, we will give a detailed
discussion of the neutrino flux produced by necklaces, a candidate
$m=3$ source.  Furthermore, we have investigated the possibility that
$m=4$ sources, whose density is constrained by bounds on the low
energy $\gamma$-ray flux, can nevertheless produce an observable flux
of neutrinos.

We studied neutrino production using a Monte Carlo approach.  In our
simulation, the red shift at which the UHE photon is emitted and its
energy are generated randomly with weight functions $\partial
J(E_\gamma, z)/ \partial z$ and $\partial J(E_\gamma, z)/ \partial
E_\gamma$ respectively.  Here $J(E_\gamma, z)$ is the differential
photon flux generated by the sources under consideration.  The red
shift dependence of the photon flux is due to the time evolution of
the sources, given by eq.~(\ref{z_dep}), and the expansion of the
universe.  The photon spectrum $\partial J(E_\gamma, z)/ \partial
E_\gamma$ is a sharply falling function of energy.  It has been
calculated for various topological defects and relic particles
in~\cite{TD_ber,pj,ps}.

Next, we let the UHE photons scatter with CMB photons, generating the
scattering angle and the energy of the background photons randomly for
each interaction.  Radio background and intergalactic magnetic fields
are insignificant for $z>z_{\rm min}$, and their effects are
neglected. The leading particle, i.e. the particle with the highest
energy produced in each interaction, is followed throughout the
electromagnetic cascade until its energy becomes too low to produce
muons.  The probabilities for competing reactions are determined by
the cross sections; which reaction actually occurs is once again
determined randomly. The reactions taken into account are $\gamma
\gamma_{_{CMB}} \rightarrow e^+ e^-$ (PP), $\gamma \gamma_{_{CMB}}
\rightarrow e^+ e^- e^+ e^-$ (DPP), $\gamma \gamma_{_{CMB}}
\rightarrow \mu^+ \mu^-$ (MP), $e \gamma_{_{CMB}} \rightarrow e \gamma
$ (ICS), $e \gamma_{_{CMB}} \rightarrow e\, e^+ e^-$ (TPP) and $e
\gamma_{_{CMB}} \rightarrow e\, \mu^+ \mu^-$ (MPP).  The cross section
for pion production is small~\cite{pi}, and neglected.

For center of mass energy $s \gg m_e^2$, the cross sections for PP and
ICS are related by $\sigma_{_{\rm PP}} = 2 \sigma_{_{\rm ICS}}$, with
$\sigma_{_{\rm PP}}$ given by eq.~(\ref{sig_PP}).  The elasticity of
these reactions is small: $\eta =1 - 2 (m^2/s)$.  For $s \lesssim 100
m^2$, the high energy approximation is not valid, and one has to
resort to the exact (tree level) expressions as can be found in any
textbook~\cite{peskin}.  The corresponding results for MP can be
obtained through the replacement $m_e \leftrightarrow m_\mu$.  The DPP
cross section quickly approaches its asymptotic value $\sigma_{\rm
DPP} \simeq 6.45 \mu{\rm barn}$.  The kinematics of this reaction has
not been calculated.  We have assumed that the energy of the incoming
photon is evenly shared by the produced particles, {\it i.e.} four UHE
electrons are produced, each with average energy $E_e = E_\gamma/4$.
This assumption does not affect the outcome much though: using instead
that only two of the electrons get all the energy yields a flux that
is almost indistinguishable from the flux obtained by the
four-electron-assumption.

The most important interactions for neutrino generation are TPP and
MPP: more than $95 \%$ of all cascade interactions are TPP, and MPP
governs muon production. For $s \gtrsim 100\, m_e^2$ the cross section
for TPP is well approximated by the Borsellino formula~\cite{tpp},
which in the high-energy limit reduces to eq.~(\ref{sig_TPP}).  The
elasticity of the leading electron is very small, eq~(\ref{eta_tpp}).
The cross sections for MPP can be obtained from TPP by replacing $m_e$
with $m_\mu$, and multiplying by a symmetry factor $S=2$.  For lower
center of mass energies we used the numerical results obtained
in~\cite{TD_ber}. If energetic enough, the electron produced together
with the muon pair in MPP can produce a pair of muons itself; we have
incorporated this effect in our computation.

\begin{figure}[t]
\centering
\hspace*{-5.5mm}
\leavevmode\epsfysize=6cm \epsfbox{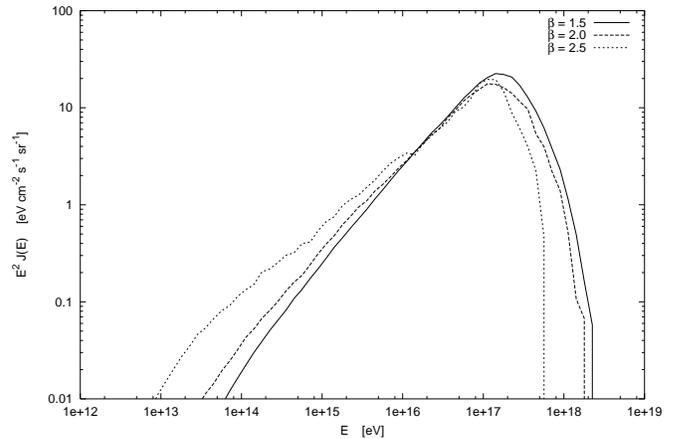}\\[3mm]
\caption[fig. 2]{\label{fig2} Differential neutrino flux $E^2 J(E)$
produced by sources with different photon spectra $n_\gamma \propto
E_\gamma^{-\beta}$: $\;\beta = 1.5, \; 2$ and $2.5$. For all plots
$m=3$, $z_{\rm max} = 10^3$, and $z_{\rm min} = 5$.}
\end{figure}

In figures \ref{fig1} and \ref{fig2} we have plotted the differential
neutrino flux produced by various photon sources.  We have assumed
that the photon emission rate can be written in the form
$\dot{n}_\gamma = \dot{n}_{0,\gamma} E_{\gamma}^{-\beta} (t/t_0)^m$,
and have varied the values of $m$ (figure~\ref{fig1}) and $\beta$
(figure~\ref{fig2}).  For all plots we have taken $z_{\rm min} =5$,
$z_{\rm max} = 10^3$ and photon energies in the range $10^{17} {\rm
eV} < E_\gamma < 5 \times 10^{20}$.  To normalize the flux, we have
assumed that the photon sources are the origin of the UHECR at
present, see eq.~(\ref{norm}).  The results agree well with our
estimates.  The highest energy neutrinos have energies $E_\nu \sim
10^{17} {\rm eV}$.  At these energies the differential flux is peaked;
for $m=3$ sources the peak value is $E_{\nu}^2 J(E_\nu) = E_{\nu}^2
{\rm d} \phi_\nu / {\rm d} E_\nu \sim 10 \, {\rm eV} {\rm cm}^{-2}
{\rm s}^{-1} {\rm sr}^{-1}$.  The neutrino flux is smaller for lower
values of $m$, and the peak value decreases accordingly: $(E^2 J)_{\rm
peak} \sim 1 - 0.1 \, {\rm eV} {\rm cm}^{-2} {\rm s}^{-1} {\rm
sr}^{-1} $ for $m=2$ to $(E^2 J)_{\rm peak} \sim 10^{-3}-10^{-2} \,
{\rm eV} {\rm cm}^{-2} {\rm s}^{-1} {\rm sr}^{-1}$ for $m=0$.  The
flux drops sharply at neutrino energies $E_\nu \sim 10^{18} {\rm eV}$
just above the peak energies, due to the cutoff at $z = z_{\rm min} =
5$.  In reality, the radio background and intergalactic magnetic field
are turned on gradually, and the cutoff is somewhat less sharp.  The
shape of the photon spectrum is not crucial for the results.  Values
of $\beta =1.5, \; 2, \; 2.5$ all give similar neutrino spectra, with
peak values differing by less than a factor $\sim 2$.

\begin{figure}[t]
\centering
\hspace*{-5.5mm}
\leavevmode\epsfysize=6cm \epsfbox{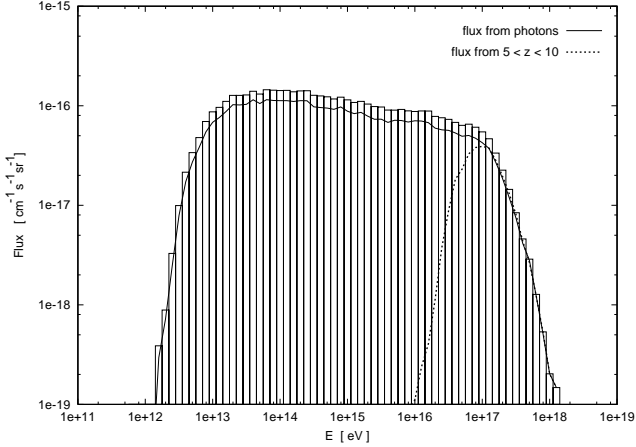}\\[3mm]
\caption[fig. 3]{\label{fig3} Neutrino $\phi_\nu$ flux produced by
necklaces.  Area under the solid line is the contribution from primary
photons to the flux.  Area under the dashed line is the contribution
to the flux from red shifts $5 < z < 10$.  }
\end{figure}

We will now discuss in more detail the results for one of the
candidate $m=3$ sources, namely necklaces.  The UHE photons emitted by
necklaces and other topological defects are the result of decaying
superheavy ``X'' particles. For example, X particles may be injected
by super conducting strings, emitted from cusps or intersections of
ordinary strings, or produced in the annihilation of monopole
anti-monopole pair~\cite{TD_review,TD,TD_ber}.  A general feature of
decaying supermassive X particles is a predominance of pions among the
decay products.  Charged pions decay through the chain $\pi \to \mu
\bar{\nu}_\mu \to e \bar{\nu}_e \nu_\mu \bar{\nu}_\mu$, producing
electrons and neutrinos.  This gives a flux of primary neutrinos,
peaked at high energies: $E_\nu \sim 10^{20} {\rm eV}$.  The photons
emitted by neutral pion decay ($\pi^0 \to \gamma \gamma$) together
with the electrons from charged pion decay also produce a flux of
neutrinos, through the mechanism described in
section~\ref{production}.  In our calculation, we have taken into
account both the UHE photons and electrons from pion decay.  Since the
average electron energy is lower than the average photon energy, the
UHE photons will give the dominant contribution to the secondary
neutrino flux.

\begin{figure}[t]
\centering
\hspace*{-5.5mm}
\leavevmode\epsfysize=6cm \epsfbox{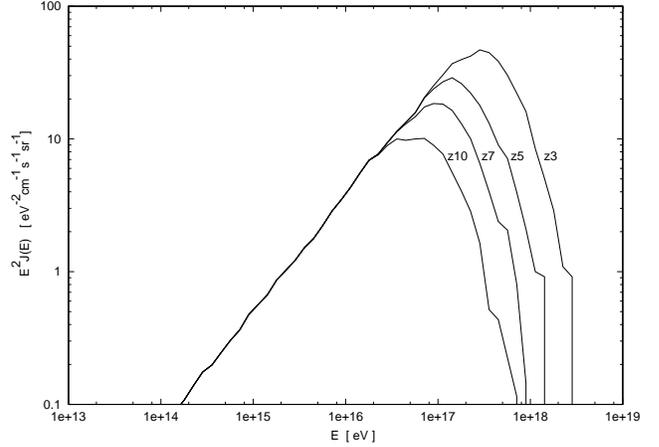}\\[3mm]
\caption[fig. 4]{\label{fig4} Differential neutrino flux $E^2 J(E)$
produced by necklaces for various values of $z_{\rm min}$. Here $z3$
corresponds to $z_{\rm min} = 3$, $z5$ to $z_{\rm min} = 5$, $z7$ to
$z_{\rm min} = 7$, and $z10$ to $z_{\rm min} = 10$.}
\end{figure}

The results for necklaces are shown in figures~\ref{fig3}
and~\ref{fig4}. In fig.~\ref{fig3} the neutrino flux as function of
neutrino energy is plotted, for $z_{\rm min} =5$, $z_{\rm max} = 10^3$
and photon energies $10^{17} {\rm eV} < e_\gamma < 5 \times 10^{20}
{\rm eV}$.  We used the photons spectrum calculated (and normalized to
fit the UHECR data) in~\cite{TD_ber}, extrapolated down to energies as
low as $E_\gamma = 10^{17} {\rm eV}$.  The contribution from the
primary photons to the neutrino flux is indicated by the solid line in
the plot.  As anticipated, it constitutes the main contribution.  The
neutrino flux generated at small red shift $5 < z < 10$ is given by
the dashed line. The total flux coming from these red shifts is of the
order $\phi_\nu \sim 10^{-16} {\rm cm}^{-2} {\rm s}^{-1}{\rm
sr}^{-1}$, in agreement with our estimate, eq.~(\ref{flux_td}).

The neutrino production mechanism is robust, and the produced flux is
not much affected by small changes in cross sections or elasticities.
Changes of $10\%$ in the MPP cross section, the elasticity of the
leading particle and of the produced muons do not change the produced
flux by more than a factor $\sim 2$.  The shape of the photon spectrum
gives an uncertainty of the same order or magnitude, as can be seen
from fig.~\ref{fig2}.  Changing the value of $z_{\rm min}$ makes a
more appreciable difference in the produced flux of highest energy
neutrinos. Fig.~\ref{fig4} shows $E^2 J(E)$ for values of $z_{\min}
=3, \;5, \;7$ and $10$.  Still, the difference in flux with cutoff
$z_{\rm min} =3$ and cutoff $z_{\rm min} = 10$ is less than a factor
$10$.  For all cutoffs the flux has peak value $E^2 J(E) =
\mathcal{O}(10)\; {\rm eV} {\rm cm}^{-2} {\rm s}^{-1} {\rm sr}^{-1}$
at the high end of the neutrino spectrum $E_\nu \sim 10^{17} - 10^{18}
{\rm eV}$.

\begin{figure}[t]
\centering
\hspace*{-5.5mm}
\leavevmode\epsfysize=6cm \epsfbox{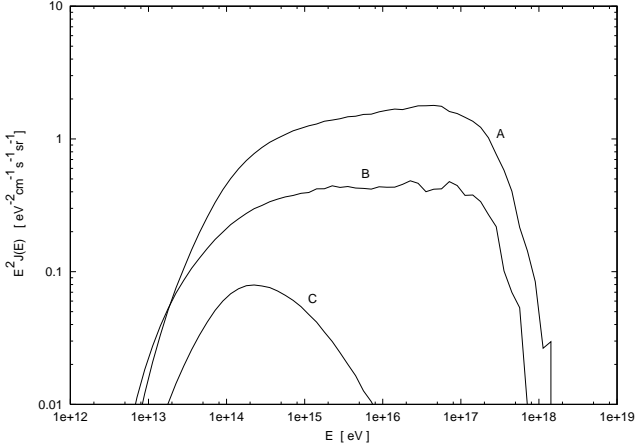}\\[3mm]
\caption[fig. 5]{\label{fig5} Maximum differential neutrino flux $E^2
J(E)$ that can be produced by $m \ge 4$ sources, in accordance with
the EGRET bound.  Line A corresponds to $m = 4$ and $\beta = 1.5$
photon spectrum, line B to $m = 4$ and $\beta = 2$, and line C to $m =
5$ and $\beta = 2.5$.}
\end{figure}

The density of $m \ge 4$ sources, such as superconducting strings, is
constrained by the EGRET bound.  The flux of the cascade photons in
the energy range $10 \, {\rm MeV} < E <100 \, {\rm GeV}$ must be lower
than the flux measured: $\omega_{\rm cas} \le 2.6 \times 10^{-6} {\rm
eV} {\rm cm}^3$.  One can estimate the flux of cascade photons
produced by TD's by assuming that all the energy density emitted
through UHE photons is transferred to the low energy photons.  The
universe becomes transparant for low energy photons at red shift $z
\sim 10^3$.  The EGRET bound then constrains the photon flux to be
\begin{equation}
\int_{z=0}^{z_{_{\rm GZK,\gamma}}} {\rm d} z \; \partial J(E_\gamma,
 z) / \partial E_\gamma \lesssim (10^{-2}-10^{-3}) \phi_{_{\rm CR}}.
\end{equation}
The maximum flux of neutrinos that can be produced by $m \ge 4$
sources when this bound is taken into account is $E^2J(E) \sim
\mathcal{O}(1)$, as shown in figure~\ref{fig3}.  Here we used $z_{\rm
min} =5$, $z_{\rm max} = 10^3$ and photon energies in the range
$10^{17} {\rm eV} < e_\gamma < 5 \times 10^{20} {\rm eV}$.  The photon
spectrum is taken to be of the form $\dot{n}_{\gamma, 0} \propto
E_\gamma^{-\beta}$. Flux A corresponds to $m=4$ and $\beta = 1.5$,
flux B to $m=4$ and $\beta = 2$.  For $m \ge 4$ sources the EGRET
constraint is more severe.  Flux C shows the result for $m=5$ and
$\beta = 1.5$.

\section{Discussion}
\label{disc}

Neutrinos from slowly decaying relic particles or some other $m \le 2$
source are too sparse to be detectable in the foreseeable future.
However, the flux from $m=3$ sources, such as necklaces, may be
detected soon.  The neutrino spectrum generated by $m=3$ sources has a
peak value $(E^2 J)_{\rm peak} = \mathcal{O}(10) \, {\rm eV} {\rm
cm}^{-2} {\rm s}^{-1} {\rm sr}^{-1}$ for neutrino energies $E_\nu \sim
10^{17} - 10^{18} {\rm eV}$.  In all our calculations we assumed that
the photon sources in question are resposible for the UHECR today.
Certainly, the flux of photons emitted cannot exceed the observed
UHECR flux, and thus our calculations can be interpreted as an upper
bound on the flux of secondary neutrinos produced.

Can other sources produce a comparable flux of neutrinos at
$10^{17}-10^{18}$eV?  The neutrino flux $ \phi_{\nu} \sim 10^{-16}
{\rm cm}^{-2} {\rm s}^{-1}{\rm sr}^{-1}$ at $E_\nu \sim 10^{17}$eV
exceeds the background flux from the atmosphere and from pion
photoproduction on CMBR at this energy~\cite{hs,stecker,sp}, as well
as the fluxes predicted by a number of
models~\cite{whitepaper}. Models of active galactic nuclei (AGN) have
predicted a similar flux of neutrinos at these energies~\cite{mpr}.
The predictions of these models have been a subject of
debate~\cite{wb}.  However, the neutrinos produced by photon sources
at high red shift have a distinctive feature: they create a sharp
``bump'' in the spectrum.  Moreover, if the photon sources are TD's
the secondary flux is accompanied by a primary neutrino flux peaked at
$E_\nu\sim 10^{20}$eV.  And everyone agrees that AGN cannot produce
neutrinos with energies of $10^{20}$eV~\cite{rm}.  So, an observation
of $10^{17}$eV neutrinos accompanied by a comparable flux of
$10^{20}$eV neutrinos would be a signature of a TD rather than an AGN.

The density of TD's with $m \ge 4$, {\em e.g.} superconducting
strings, is constrained by the EGRET bound on the flux of low energy
$\gamma$-photons.  This also constrains the neutrino flux these
sources can produce to be less than $E^2 J(E) \le \mathcal{O}(1) \,
{\rm eV} {\rm cm}^{-2} {\rm s}^{-1} {\rm sr}^{-1}$.  This is probably
too low to be detected.

To summarize, we have shown that sources of ultrahigh energy photons
that operate at red shift $z \gtrsim 5$ produce neutrinos with energy
$E_\nu \sim 10^{17}$eV.  The flux depends on the evolution index $m$
of the source.  A distinctive characteristic of this type of neutrino
background is a ``bump'' in the spectrum at neutrino energies $E_\nu =
10^{17}-10^{18}$eV, and a sharp cutoff at $E_\nu \sim 10^{18}$eV due
to the universal radio background at $z<z_{\rm min}$.  We have
calculated numerically the neutrino spectrum produced by various
sources.  The produced flux is largest for $m=3$ sources. It has a
peak value $E^2 J(E) = \mathcal{O}(10)\, {\rm eV} {\rm cm}^{-2} {\rm
s}^{-1} {\rm sr}^{-1}$ at neutrino energies $E_\nu \sim 10^{17} -
10^{18} {\rm eV}$.  Detection of these neutrinos can help understand
the origin of ultrahigh energy cosmic rays.  Sources with larger
values of the evolution index $m \ge 4$ are ruled out as the origin of
observed UHECR by the EGRET bound on the flux of $\gamma$-photons.
This bound also constrains the flux from these sources to be less than
$(E^2 J)_{\rm peak} \le \mathcal{O}(1)\, {\rm eV} {\rm cm}^{-2} {\rm
s}^{-1} {\rm sr}^{-1}$.

The author would like to thank A.~Kusenko and C.~Postma for useful
discussions. This work was supported in part by the US Department of
Energy grant DE-FG03-91ER40662, Task C, and by a grant from UCLA
Council on Research.


\end{document}